# Case Study for Developing a UXR Point of View for FinOps Product Innovation

Jason Dong
Google
Kirkland, WA, US
jasondong@google.com

Anna Wu
Google
Kirkland, WA, US
annwu@google.com

## ABSTRACT

In the dynamic landscape of Cloud financial management, we are sharing a case study exploring the development of a User Experience Research (UXR) Point of View (PoV) to drive FinOps product innovation. We demonstrate how qualitative and quantitative research methods working together to navigate the challenges of understanding customer needs, aligning cross-functional teams, and prioritizing limited resources. Through a multi-phased research approach, the research team identifies opportunities, quantifies pain points, and segments diverse customer cohorts. This culminated in a UXR PoV that informed the creation of a differentiated product strategy, a 'one-stop shop' dashboard empowering FinOps practitioners with actionable insights and tools. This case study highlights the power of mixed-methods research in uncovering actionable insights that drive impactful product innovation.

## 1 Background

A few years back, Google was developing a new cloud product to help customers manage their cloud spending. The product team entered the market of FinOps (Financial Operations) [1] with a few established competitors. The product team aimed to innovate and disrupt the space with unique offerings. Due to engineering resource constraints, the product team needed to be highly selective about where they invest.

We believe this is a representative scenario, as cost optimization is a common customer pain (a necessity for survival, not a "nice to have") and developing solutions facing triple constraint: time, resource and innovation. Thus, the rest of the proposal will focus on a case study with discussion points as extension.

### 1.1 Challenges

The most pressing challenge for Google's product team was to find a unique differentiator that will allow them to show thought leadership and surpass the competition even in a specific area. The product team needed a comprehensive understanding of the competitive landscape, user needs, and prevailing industry trends to inform a robust product strategy. These three were among the most challenges specifically for this use case example:

1. Challenge #1: Despite the necessity and desirability, understanding and optimizing Cloud spend are fairly new to customers as well as to cloud providers. Customers can over focus on migration and functionality, leading to unexpected cost; understanding a wide range of new pricing models (metered, subscription, reservation…) and choosing the right one for their workload is hard; with fast evolving cloud technology (microservices, containers), the increasing dynamic and complex architecture introduces new challenges in tracking usage and managing cost.

2. Challenge #2: Focus on high-impact customer needs given the triple constraints. Diverse customer voices raised through different channels (sales, forums, in-product feedback, user research). Some are louder than others. Like many early businesses, current revenue distribution is highly left-skewed, the product team needs to come up with a weighted way to integrate customer feedback.

3. Challenge #3: No clear market sizing for the proposed solutions. Given its evolving nature, diverse customer needs and limited data availability (scarcity and confidentiality), predicting potential markets is like a moving target. Yet, being able to estimate is critical to get leadership buyin and thus resource prioritization.

# 2 The Research POV

To effectively address these multifaceted challenges, the research team worked with cross-functional partners clarifying business objectives and developed multi-phased research plans, involving both qualitative and quantitative research methods that guided product development efforts. The pyramid framework [2] provided by User Experience Research Playbook helps provide a structure for illustrating the example.

## 2.1 Foundational

In the rapidly innovating FinOps domain, developing a differentiating feature from a catch-up position is a major hurdle. To set clear business objectives and define research project's purpose, a series of cross-functional workshops were held with UXR, UXD, PM, and Engineering teams to align on the existing knowledge about competitive landscape, who are the potential customers, their needs, and product gaps. These activities helped align cross functional teams to understand FinOps market, define product vision, and identify gaps(e.g. research questions) needed to help them get to where they want to be. In parallel and informed by these workshops, The research team conducted literature review of existing knowledge (including persona, CUJ, and historical customer pain points), crafted a thorough competitive analysis proposal, planned for evaluative survey as well as logs analysis to address the opportunity sizing questions when we have a proposal.

This phase presented several key challenges: discerning distinct customer needs by archetypes across diverse channels, managing escalations from vocal customers while identifying core issues, and achieving a shared understanding of scope and business target. In large corporations, early alignment on research and target customers is often overlooked due to distributed teams and competing priorities. However, getting to the root of the research request and accessing the business risk the requestor is facing is essential to the success of the research project and product development. Common concerns around engaging researchers(and each other) could include:

- creating more challenges and slowing them down;
- study information could be inconclusive and confuse them more;
- study information could conflict with current assumptions.

To address these concerns, we shifted our approach to research planning. Collectively, we developed a proposal to reduce uncertainty and establish a robust and consensus-driven understanding, recognizing its critical importance for subsequent activities.

## 2.2 Data Collection

Recent UX research literature [3] indicates a surge in mixed methods approaches, enhancing the discipline's ability to demonstrate its value and address diverse research needs. To generate representative and meaningful insights, a few rounds data collection with various research methodologies were conducted:

- The competitive analysis aims to understand the existing market, identify gaps, and find differentiators by reviewing documentation, analyzing competitor strategies, identifying industry trends, and engaging consulting and analyst firms. Research activities included 30+ interviews, evaluations of 10+ user journeys, and mark user pain points based on severity. The competitive analysis was comprehensive qualitative research with rich data showing why certain types of customers need to have certain pain points addressed. It showed where we are behind and how much further competitors are in their offerings, initially identified the opportunities for product differentiation.

- Building on this qualitative foundation, we conducted quantitative research through large-scale surveys and log analysis. This is critical to figure out the opportunity sizing that is critical for product teams' decision making in prioritization. The quantitative approach allowed us to segment users, classify and sizing pain points across cohorts(e.g. SMBs, digital natives, traditional enterprises), quantify diverse needs and value prop(revenue impact). For example, we determined X% of our target customer base are novice users, requiring UI enhancements versus advanced users needing programmatic access. This data helped contextualize those feature requests from vocal minorities (e.g. enterprise customer X asked for a migration feature, escalated through the accounting team to leadership and ended up an outlier) and informed market sizing for the product team. Triangulated from altitudinal and behavioral data boosted stakeholder confidence that eventually led to product funding.

Themes in this case also recurring elsewhere, e.g. the sentiment of "User Feedback is Overrated". As qualitative researchers, we strive to elevate the customer and user perspective in product development. It's disheartening when research requests are framed solely for validation, demanding numerical evidence. Even more concerning is the tendency to dismiss self-reported data in favor of large-scale behavioral metrics like api requests, dwell time, or navigation paths, particularly when confronted with smaller sample size. However, understanding the true intent of cross-functional requests is crucial. Often, the requester's needs are addressed by qualitative insights, requiring a deeper understanding to inform decisions. Conversely, if the request centers on sizing and requires numerical evidence, quantitative methods are essential. In many cases, a combination of both is ideal. In our case, we employed a sequential exploratory approach: qualitative research to frame the problem, followed by quantitative research to size the opportunity. In other situations, an interpretive sequence—quantitative research shaping the pattern followed by qualitative research for explanation (e.g. post-survey interview to identify rationales behind negative sentiment trending)—may be more informative. Additionally (more likely when teams are distributed), a parallel approach, cross-validating independently collected data, is required to significantly enhance confidence.

## 2.3 Insight Generation

The competitive analysis provided a wealth of qualitative data on customer pain points and competitor performance, resulting in a lengthy list of opportunities. To refine this list, the product team applied heuristics, evaluating each opportunity based on feasibility, ownership, competitive landscape, and customer impact. A list we used to prioritize:

- Do we have confidence to fix an issue within a reasonable time frame?
- Whether the product team owns the problem end to end?
- Are competitors already leading in the area?
- How critical is the problem to the customer's core FinOps task?

While this process reduced the number of potential focus areas, selecting a truly differentiating feature to achieve thought leadership still proved difficult within resource limitations.

As previously stated, our quantitative survey and log analysis provided critical sizing estimates for each pain point, segmented by user cohort and company archetype. Hundreds of customer survey responses, coupled with formagraphic data for clustering, allowed us to quantify the severity of these pain points. Log analysis further enriched our understanding by linking billing patterns to revenue and assessing feature reach. We identified specific surfaces through billing features with higher headroom for growth. A refined list of opportunities emerged, revealing key insights. For example, we discovered that FinOps practitioners, as pioneers in this rapidly developing discipline, seek opinionated guidance and best practices from cloud providers to direct their energy and efforts. This ultimately led to the development of POV, a concept of a 'one-stop shop' for monitoring, learning, and performing nuanced tasks. The quantitative analysis suggested the significant market opportunity, especially with low to mid-FinOps mature customers and also found the potential for high FinOps mature customers to compare their current practices with what is available in the product. Further resource assessments confirmed the implementation feasibility.

Inevitably, we encountered conflicting data points (attributed by the diverse customer cohorts). We believe in leaning into these conflicts, never leaving them unexplained. We actively seek out these moments of divergence, recognizing them as opportunities for deeper understanding and refined insights. This goes back to "Data is the Compass, Not the Destination." While numbers provide essential guidance, they should not stifle creativity. It is much easier to simply present data and leave interpretation to others, which risks confusion and potentially reluctance in engaging research when the team is facing constraints. Transforming complex data into compelling narratives is essential for driving action and ensuring insights translate into truly innovative and delightful experiences.

## 2.4 UXR PoV

This process culminated in a compelling UXR Point of View (PoV) to drive impact. Facing triple constraints, we aligned cross-functional commitment to a consolidated dashboard concept. The dashboard, envisioned as a 'one-stop shop' for FinOps practitioners, aims to provide opinionated guidance and facilitate monitoring and task execution, addressing critical pain points by different cohorts. This concept was eventually launched as a portal catering to both daily operations and educational needs for many Cloud customers. Recognizing the inherent uncertainty in product development, we adopted a 'strong opinions, loosely held' approach. We leveraged intuition, multiple research methodologies, and synthesis to inform teams, acknowledging its potential for evolution. This iterative process, involving stakeholder feedback and testing through rounds of research, allows us to adapt, learn, and refine our perspective by triangulating various sources.

## 3 Conclusion and Discussion

This use case illustrates our journey through the four levels of the UXR POV pyramid. Facing triple constraints, we collaborated with cross-functional teams, adopting multiple research methods, distilled insights that eventually led to creating a differentiated, user-centered, and successful product in the FinOps space. This approach facilitates a clear and actionable path for product development, grounded in a deep understanding of user needs and the competitive market.

With the pressing AI technology, we would love to extend the discussion with workshop participants around the distinct value of human intuition for UX research practitioners, especially going beyond the self-labeled "quant UX researcher" or "qual UX researcher". Bold predictions like "The Future of UX Research is AI-Driven" [4] suggests that artificial intelligence will eventually surpass human capabilities in analyzing user data and generating insights. The past couple years' reality also proved that practitioners increasingly start to rely on AI tools [5]. This emergent practice once again emphasizes developing a well-rounded POV is essential for credibility and persuasive communication. A lack of quantitative knowledge limits our ability to size opportunities, understand user behavior at scale, and validate qualitative findings. Conversely, a lack of qualitative understanding obscures the nuances of user experience, leading to incomplete or misleading interpretations. Be it a case study that creates a personal connection from the unmet needs found in thousands of consumer reports, or a default setting that benefits majority strengthen with social commentary, Our unique humanity strength resides when we embrace embrace both qualitative and quantitative methods as essential tools in our research toolkit. Developing proficiency in both is to our distinct advantage, and a mindset shift is required to achieve this. Quant is often perceived as challenging, but tools like generative AI can facilitate learning, script development, and analysis. Furthermore, collaborating with data professionals ensures a more comprehensive and robust research process. We are not going to be replaced by AI, but someone who uses AI.